\documentclass[a4paper,12pt]{article}
\usepackage{marvosym}
\usepackage{comment}
\usepackage{amsmath, amssymb}
\usepackage{bbold}
\usepackage{subfigure}
\usepackage{mathrsfs}
\usepackage[framemethod=default]{mdframed}
\usepackage{tikz}
\usetikzlibrary{decorations.pathmorphing}
\usepackage{microtype}
\usepackage{booktabs}
\usepackage{enumitem}
\usepackage{caption}
\usepackage{siunitx}
\usepackage[dvipsnames]{xcolor}
\usetikzlibrary{arrows.meta}
\definecolor{UrlGarnet}{HTML}{A33B4A}

\newif\ifeditfinal \editfinalfalse

\newmdenv[skipabove=1pt,
skipbelow=1pt,
rightline=false,
leftline=false,
topline=false,
bottomline=false,
backgroundcolor=gray!10,
linecolor=gray,
innerleftmargin=5pt,
innerrightmargin=5pt,
innertopmargin=-10pt,
innerbottommargin=5pt,
leftmargin=0cm,
rightmargin=0cm,
linewidth=4pt]{eBox}

\newcommand{\T}{^{\mathsf{T}}}
\setlist[itemize]{leftmargin=1.5em, itemsep=2pt, topsep=3pt}
\setlist[enumerate]{leftmargin=1.7em, itemsep=2pt, topsep=3pt}
\makeatletter
\providecommand{\ab}{}\providecommand{\sb}{}
\renewcommand{\ab}[2]{\mathopen{\langle}#1#2\mathclose{\rangle}}
\renewcommand{\sb}[2]{\mathopen{[}#1#2\mathclose{]}}
\makeatother

\newcommand{\PTden}{\ab{1}{\sigma(2)}\,\ab{\sigma(2)}{\sigma(3)}\,\ab{\sigma(3)}{\sigma(4)}\,\ab{\sigma(4)}{5}\,\ab{5}{1}}
\newcommand{\PTdens}{\sb{1}{\sigma(2)}\,\sb{\sigma(2)}{\sigma(3)}\,\sb{\sigma(3)}{\sigma(4)}\,\sb{\sigma(4)}{5}\,\sb{5}{1}}
\newcommand{\fff}{f^{a_1a_{\sigma(2)}b_1}f^{b_1a_{\sigma(3)}b_2}f^{b_2a_{\sigma(4)}a_5}}

\newcommand{\Ssum}{\mathcal{S}}
\newcommand{\Sbar}{\overline{\mathcal{S}}}

\catcode`\,12

\pdfoutput=1
\usepackage{jheppub}
\usepackage[T1]{fontenc}

\definecolor{ElectricIndigo}{HTML}{4F46E5}
\definecolor{HotMagenta}{HTML}{C026D3}
\definecolor{CyberTeal}{HTML}{0891B2}

\definecolor{LinkSlate}{HTML}{2B5070}
\definecolor{CiteSlate}{HTML}{395B7A}
\definecolor{UrlBrown}{HTML}{6B3A2E}
\hypersetup{
  colorlinks=true,
  linkcolor=LinkSlate,
  citecolor=CiteSlate,
  urlcolor=UrlBrown,
  filecolor=LinkSlate,
  breaklinks=true
}

{\title{Magic without a phase: phase-independent stabilizer R\'enyi entropy in gluon scattering}}

\author{Jinwei Chu, Savan Kharel}

\affiliation{Department of Physics, University of Chicago, Chicago, IL 60637, USA}

\emailAdd{jinweichu@uchicago.edu}
\emailAdd{skharel@uchicago.edu}

\abstract{
Magic, also known as non-stabilizerness, measures the usefulness of a quantum state for quantum computation. While magic is defined relative to a choice of computational basis, in some physical settings the available data determine this basis only up to local phase conventions. In this paper, we generalize the notion of magic and formulate it in a phase-independent manner, and hence define a generalized stabilizer R\'enyi entropy. As a case study, we consider higher-multiplicity tree-level gluon scattering, interpreting the outgoing helicities as qubits. In this setting, the helicity data naturally determine a local basis for each qubit but leave a phase ambiguity. For $3\to 2$ scattering, we find that the final-state phase-independent magic is generically larger than the maximum attainable in $2\to 2$ scattering. For $2 \to 3$ scattering, we find a nonzero minimal value approached in the soft limit. Moreover, when the three outgoing momenta become symmetric, the magic approaches a local minimum only a few percent above the soft-limit value. In all cases considered, the color dependence cancels from the phase-independent stabilizer Rényi entropy.
}

\begin{document} 
\maketitle
\flushbottom
\section{Introduction}
One of the salient lessons of quantum computing is that entanglement,
however elaborate, does not by itself mark the boundary of efficient classical
simulation. The Gottesman--Knill theorem makes this
point~\cite{Gottesman:1998hu,Aaronson:2004xuh,Nielsen:2012yss}: Clifford circuits
acting on stabilizer states may generate extensive entanglement and yet remain
efficiently simulable by classical computers. Every Bell state is maximally entangled; every Bell state
is nevertheless a stabilizer state.

What efficient classical simulation cannot easily reproduce is
\emph{magic}.\footnote{The terminology descends from the ``magic states'' of
Bravyi--Kitaev distillation~\cite{Bravyi:2004isx}: the non-stabilizer states
that, injected into an otherwise Clifford circuit, lift it to universality. The
name has interestingly become part of the lore.} So what is magic?
Simply put, it is how far a quantum state lies from the stabilizer
states~\cite{Bravyi:2004isx,Emerson:2013zse,Howard:2014zwm,Beverland:2019jej,Leone:2021rzd}. It
is magic, not entanglement alone, that supplies the operative resource behind
quantum advantage, the quantity distilled in stabilizer protocols and consumed
by fault-tolerant quantum computation~\cite{Leone:2021rzd,Haug:2023hcs,Leone:2024lfr}. To
ask whether the output of a physical process is genuinely hard to classically
simulate is to ask about its magic.\footnote{Recent results make this operationally sharp: ensembles of states with
little magic can be computationally indistinguishable from highly magical
ones~\cite{Gu:2023qqq}, and the very tractability of entanglement is itself
controlled by magic~\cite{Gu:2024qvn}.}

A state low in magic remains close to the stabilizer world. The stabilizer
R\'enyi entropies~\cite{Leone:2021rzd}, now established as genuine resource
monotones for R\'enyi index $q \geq 2$~\cite{Haug:2023hcs, Leone:2024lfr},
make this computable. It expresses magic through Pauli expectation values and turns it into a tractable observable across many-body
physics~\cite{Haug:2022vbl,Oliviero:2022euv,
Haug:2022vpg,Odavic:2022wfn,Lami:2023naw,Haug:2023ffp,
Lami:2024osd,Lopez:2024jjq,Dowling:2024wvo,Ding:2025nua,Jasser:2025myz,Bera:2025pfp,
Falcao:2025vvk,Tirrito:2025ocm,Hallam:2026xvn,Bettaque:2026vpl},
conformal and quantum field theory~\cite{White:2024nuc,White:2024bjp,Chernyshev:2024pqy,Cepollaro:2024sod, Liu:2025qfl,Hoshino:2025jko,Hoshino:2025ine,Gargalionis:2025iqs,Liu:2025bgw,Viaux:2026skf,Cao:2026aye,
Matsuda:2026ttw}, and
lattice gauge theory~\cite{Jha:2026ror,Santra:2025dsm,Gupta:2026hif,
Crew:2025khs}.

Yet whether a state is a stabilizer state or not depends on a choice of basis. Consequently, nonstabilizerness, and thus all magic measures including the stabilizer R\'enyi entropy, is not, by itself, an invariant of the physical state. These measures are invariant under Clifford transformations by construction,
but not under generic local unitary rotations. In particular, a local
rephasing of the qubit basis changes the assigned magic continuously: even a state that is a Bell state, and hence a canonical zero-magic stabilizer state, in one basis can acquire a continuous range of nonzero stabilizer Rényi entropies under a change of basis phase, as we review in Section~\ref{sec:phase}.

It is thus useful to generalize the notion of magic so that it is invariant under rephasings of the qubit basis. To this end, we define phase-independent analogs of the stabilizer R\'enyi entropies. This is the central
construction of the present work: a prescription for quantifying magic whenever
the computational basis is fixed only up to phase. We apply it  to a concrete case study: tree-level gluon scattering amplitudes.

A scattering amplitude is, on one poetic view, the most
perfect microscopic structure in the universe~\cite{Dixon:2011xs}; from a
quantum-information standpoint it is already a quantum object before any
probability is formed from it, a coherent superposition constrained by locality,
unitarity, gauge invariance, and Lorentz symmetry. Concretely, once helicities are retained,
the final state of an $m\to k$ gluon scattering process lives in a $k$-qubit Hilbert space, and
the amplitude is the wavefunction prepared by the
$S$-matrix~\cite{Cervera-Lierta:2017tdt,Garrido:2025xpk, Beane:2018oxh,Low:2021ufv,Fedida:2022izl,Cheung:2023hkq,Aoude:2024xpx,McGinnis:2025brt}.\footnote{The
quantum-information content of the $S$-matrix has been studied mostly through entanglement: how scattering generates
it~\cite{Seki:2014cgq,Peschanski:2016hgk} (see~\cite{Balasubramanian:2011wt} for momentum-space entanglement in QFT), how it
ties to the cross section~\cite{Cheung:2023hkq, Low:2024mrk, Low:2024hvn}, and how its suppression signals emergent
symmetry~\cite{Beane:2018oxh,Low:2021ufv, Liu:2022grf}. Magic is the
resource one rung up.}  Besides, in the scattering amplitudes community, gluons are among the sharpest probe, and few amplitudes have been studied more thoroughly. The results have been impressive: Parke--Taylor~\cite{Parke:1986gb},
twistors~\cite{Witten:2003nn,Cachazo:2004kj}, BCFW~\cite{Britto:2004ap,Britto:2005fq},
the double copy~\cite{Kawai:1985xq,Bern:2008qj,Bern:2010ue,Bern:2019prr} were all borne out of this investigation.

The study of magic in scattering amplitudes began with the top quark~\cite{White:2024nuc, White:2024bjp}, whose spin
correlations, following the proposal of~\cite{Afik:2020onf}, are now measured directly at the LHC~\cite{ATLAS:2023fsd,CMS:2025cim} as part of a
fast-growing program of quantum information at
colliders~\cite{Barr:2024djo,Guo:2026yhz}. Since then it has reached gluon
and graviton scattering~\cite{Gargalionis:2025iqs,Gargalionis:2026onv}, quantum
electrodynamics~\cite{Liu:2025qfl}, Schwinger pair production~\cite{Grieninger:2026txt}, the electroweak sector and its principle
of minimal magic~\cite{Liu:2025bgw,Cao:2026aye}, two-Higgs-doublet
models~\cite{Busoni:2025dns}, the top sector and particle decays as probes
of new physics~\cite{Aoude:2025jzc,Viaux:2026skf}, and nuclear,
hypernuclear, and neutrino
systems~\cite{Robin:2024bdz,Li:2026udy,Chernyshev:2024pqy,Froustey:2026slw,Robin:2026lqp}. Automated
tools now compute these observables for generic
processes~\cite{Durupt:2025wuk}, and measurements have been proposed at the
Electron--Ion Collider~\cite{Cheng:2025zaw,Fucilla:2026mkg}.

The picture developed in these works is compelling and has opened a useful direction. However, these studies rest on particular choices of state basis, which our
phase-independent measure now refines by quotienting out the basis-phase ambiguity. Note that the phase ambiguity can also be removed through a minimization of magic over all choices of the basis phase, as in the definition of non-local magic~\cite{Cao:2023mzo,Qian:2025oit}
(see also~\cite{Fliss:2020yrd,Robin:2025ymq,Malvimat:2026oqf,Gargalionis:2026onv,Busoni:2026lvp} 
for minimized or non-local magic of scattering amplitudes, linked states and the SYK model). In contrast, our construction involves an \emph{average} over the basis-phase $U(1)^{k}$ orbit (a typical-phase rather than a best-case invariant)\footnote{See also~\cite{White:2024nuc, CMS:2025cim,Iannotti:2025lkb} for related study of averaged magic and \cite{Turkeshi:2023lqu} for an interesting study of typical (Haar-random) states.}, and we carry it beyond $2\to2$ to the five-point ($3\to2$ and $2\to3$) amplitudes. 

We work at tree level, where the amplitudes are known exactly and the number of
outgoing gluons is a parameter we can dial. Three processes tell the story. In
$2\to2$ scattering: the phase-independent stabilizer R\'enyi entropy is a function of the scattering angle alone, vanishing at
$\theta=0$ and $\pi$ and peaking only modestly in between. We further carry the analysis of magic to higher multiplicity five point amplitudes. In $3\to2$ scattering a
third helicity channel opens and the final state turns genuinely complex. In the case we consider, where the incoming helicity configuration is nontrivial and the incoming momenta are symmetric, the
phase-independent magic never vanishes and climbs to $\ln(27/13)\approx0.731$, far
beyond anything $2\to2$ can reach. 

In $2\to3$ scattering, the only three-qubit case in this paper, we find that the extra outgoing particle buys less magic: the
phase-independent magic is trapped between a nonzero floor, approached in the soft limit, and a
ceiling that, at this same five-point multiplicity, sits well below the $3\to2$
value. Interestingly, there is another limit corresponding to the fully symmetric configuration for outgoing momenta, where the magic reaches a local minimum, that is larger than the global one (corresponding to the soft limit) by only few percentage. 

Running through all three processes is a surprise we did not put in by hand: the
color dependence cancels, and what survives is a phase-independent magic fixed by the
kinematics alone, similar to what happens for magic and nonlocal magic in $2\to 2$ gluon scattering \cite{Gargalionis:2026onv,Robin:2025ymq}. We establish this for the MHV/anti-MHV structure of the amplitudes we treat, and leave its general validity open.

Our paper is divided into two parts. The general part includes Sections \ref{sec:measure} and~\ref{sec:phase}.  Section~\ref{sec:measure} sets up the (bare) stabilizer R\'enyi
entropy and, through a handful of two-qubit states simple enough to compute by
hand, pulls magic cleanly apart from entanglement. Section~\ref{sec:phase}
locates the problem with the magic measure: it depends on a choice of basis
phase, and removes it.

The second part applies the construction to case study:  tree-level gluon scattering. Section~\ref{sec:setup} builds the dictionary that makes all of this a question
about amplitudes: a gluon's two helicities furnish a qubit, a $k$-gluon final state
is a $k$-qubit system, and the amplitude is the wavefunction the $S$-matrix prepares.
Section~\ref{sec:examples} is where the work is, the three processes above,
worked in full. We close with directions for further work.

\section{Magic, and how to measure it}
\label{sec:measure}
We begin pedagogically, both to fix conventions and to make the discussion
accessible beyond the quantum-information community. Since the stabilizer
R\'enyi entropy is less familiar outside that setting, and is easily conflated
with entanglement measures, we spell out its definition and interpretation
through a few simple examples.

Among the proposed measures of non-stabilizerness, the stabilizer R\'enyi entropy
(SRE)~\cite{Leone:2021rzd} stands out because it is computable directly from
Pauli expectation values  and has recently been
given a direct operational interpretation~\cite{Bittel:2025yhq}. It is the measure we adopt throughout this paper.
For a pure state $|\psi\rangle$ of $k$ qubits, the SRE of R\'enyi index $q$
is~\cite{Leone:2021rzd}\footnote{See~\cite{Wang:2023uog} for the generalization
beyond qubit systems.}

\begin{equation}
\label{Mq}
M_q(|\psi\rangle)
=
\frac{1}{1-q}
\ln\!\left(
\frac{1}{2^k}
\sum_{P \in \mathcal{P}_k}
\left|
\langle\psi|P|\psi\rangle
\right|^{2q}
\right),
\end{equation}
where the sum runs over the Hermitian $k$-qubit Pauli strings
$\mathcal{P}_k = \{I, \sigma_x, \sigma_y, \sigma_z\}^{\otimes k}$.

Note that the expectation value $\langle\psi|P|\psi\rangle$ is real because each
$P$ is Hermitian. The various $M_q$ are qualitatively similar. For example, they all vanish if and only if $|\psi\rangle$ is a stabilizer state. We focus
throughout on the second SRE, $M_2$, the most widely studied member of the
family. At $q=2$, Eq.~\eqref{Mq} reduces to
\begin{equation}
\label{M2def}
M_2(|\psi\rangle)
\;=\;
-\ln \left(\,\sum_{P \in \mathcal{P}_k}
\frac{\langle\psi|P|\psi\rangle^4}{2^k}\right).
\end{equation}

\subsection {A worked example} Let us consider a one-parameter family of two-qubit states
\begin{equation}
\label{psir}
|\psi\rangle_r = \frac{1}{\sqrt{1+r^2}}\bigl(r\,|{+}{-}\rangle + |{-}{+}\rangle\bigr), \qquad r \in \mathbb R,
\end{equation}
where $\{|+\rangle,|-\rangle\}$ denotes the basis for each qubit. This basis may be interpreted, for example, as the spin-up/spin-down basis for an electron, as the left-/right-handed circular polarization basis for a photon, or as the energy eigenstates of a two-level system. For the benefit of the readers, we would like to point out that the gluon scattering serves as the principal testbed for the formalism developed here: there, this same basis assumes its physical guise as helicity. 

\begin{figure}
\centering
\includegraphics[scale=0.8]{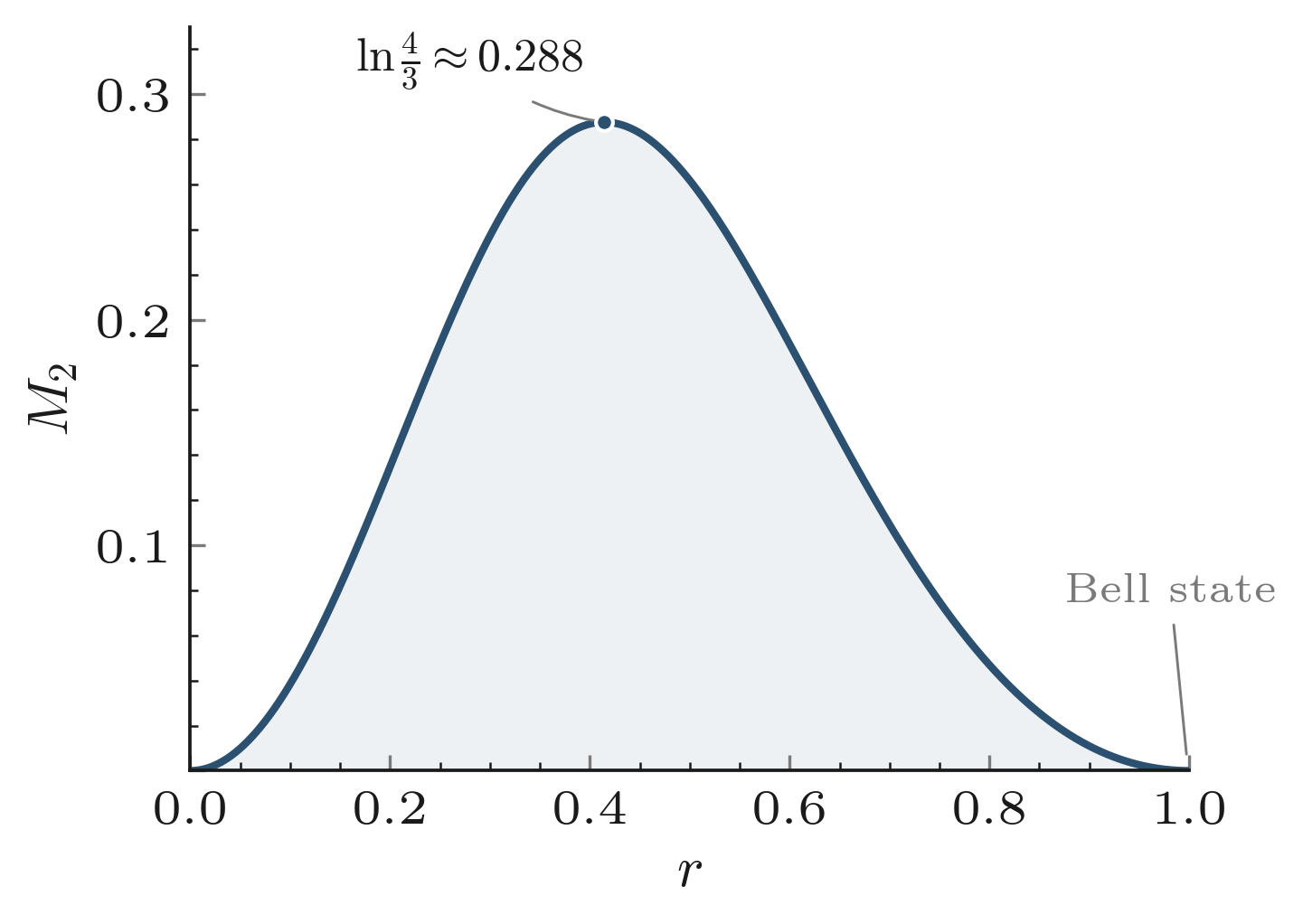}
\caption{\label{fig:M2r} The second stabilizer R\'enyi entropy $M_2$ for the two-coefficient state Eq.~\eqref{psir}, as a function of the amplitude ratio $r$. The entropy vanishes at $r=0$ (a trivial state) and at $r=1$ (the Bell state), showing that entanglement alone does not imply magic. The maximum $\ln(4/3) \approx 0.288$ is attained at $r = \sqrt{2} -1$.} 
\end{figure}

A direct calculation for (\ref{psir}) using the definition (\ref{M2def}) gives
\begin{equation}
\label{M2r}
M_2(|\psi\rangle_r) \;=\; \ln \frac{(1+r^2)^4}{1 + 14 r^4 + r^8}.
\end{equation}
Figure~\ref{fig:M2r} plots Eq.~\eqref{M2r} as a function of $r$. Due to the symmetry under $r\to -r$ or $r\to 1/r$, it is sufficient to focus on the region $r\in [0,1]$. In this fundamental region, the function vanishes at $r=0$ and $r=1$. The first zero, $|\psi\rangle_0 = |{-}{+}\rangle$, is trivial. The second is more telling. The state $|\psi\rangle_1 = (|{+}{-}\rangle + |{-}{+}\rangle)/\sqrt{2}$ is the Bell state, and is \emph{maximally} entangled; yet $M_2$ vanishes, because every Bell state is a stabilizer state. The example makes manifest that entanglement and magic are independent resources: maximal entanglement does not imply any magic at all. 

Furthermore, setting $\partial M_2/\partial r = 0$ in Eq.~\eqref{M2r} gives
\begin{equation}
\label{s0}
r(r^6 - 7r^4 + 7r^2 - 1) \;=\; r(r^2-1)(r^4 - 6r^2 + 1) \;=\; 0,
\end{equation}
with three roots $r = 0$, 1, and $ \sqrt{2}- 1$ in the region $[0,1]$. The roots $r=0$ and $r=1$ correspond to the minima with zero magic as discussed above. At $r=\sqrt{2}-1$, the magic reaches the maximum
\begin{equation}
\label{M2rmax}
M_2(|\psi\rangle_{r=\sqrt{2}- 1}) \;=\; \ln\!\frac{4}{3} \;\approx\; 0.288,
\end{equation}
the largest value attainable along the family~\eqref{psir}.

\subsection{Adding a coefficient}

\begin{figure}
\centering
\includegraphics[scale=0.8]{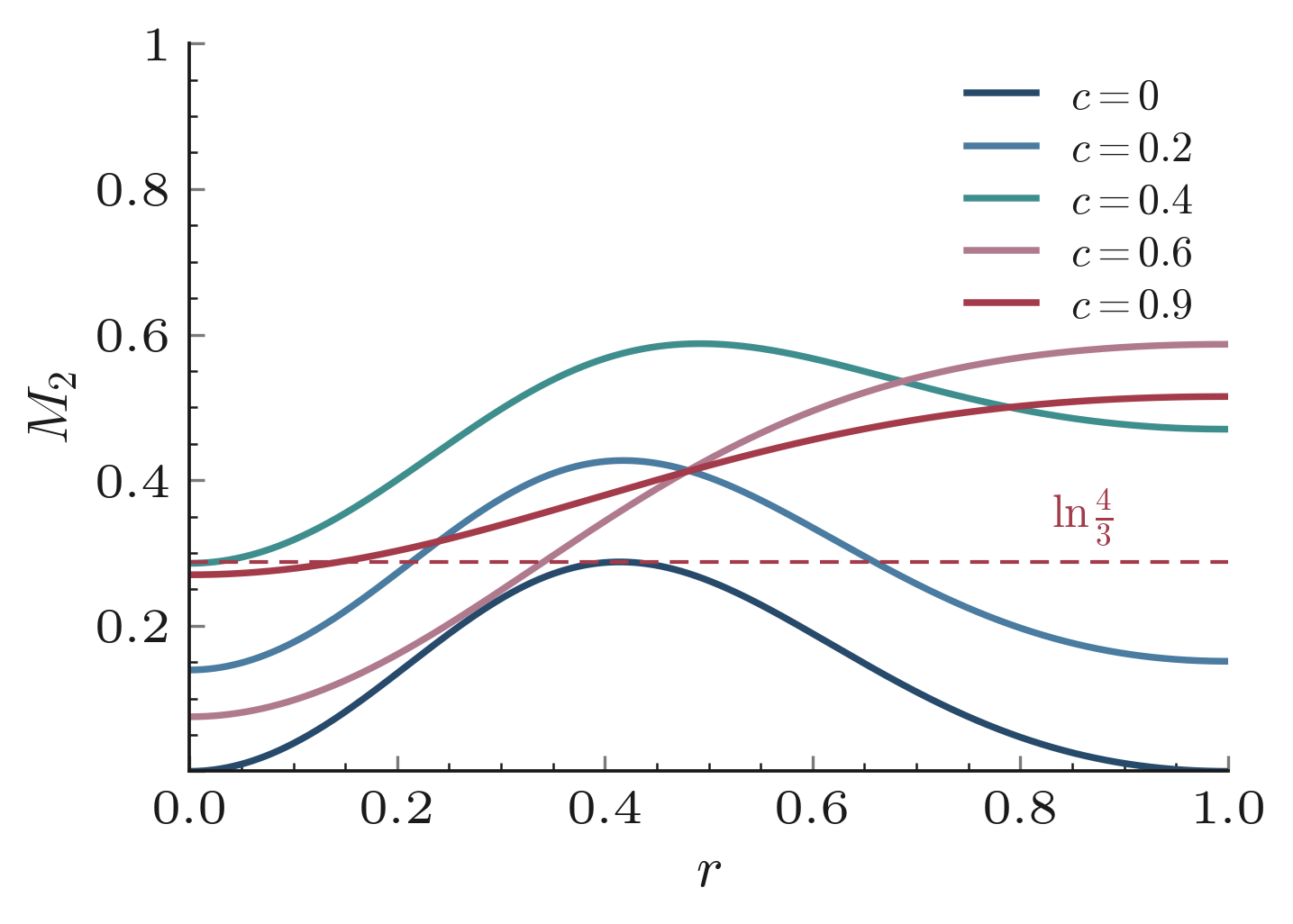}
\caption{\label{fig:M2rc} The second stabilizer R\'enyi entropy $M_2$ for the two-coefficient state Eq.~\eqref{psirc}, plotted against $r$ at several values of the admixture parameter $c$. Even a modest $c$ produces magic that comfortably exceeds the $c=0$ bound $\ln(4/3) \approx 0.288$ from Eq.~\eqref{M2rmax}.}
\end{figure}

A natural generalization of Eq.~\eqref{psir} adds a $|{-}{-}\rangle$ component,
\begin{equation}
\label{psirc}
|\psi\rangle_{r,c} \;=\; \sqrt{\frac{1-c^2}{1+r^2}}\,\bigl(r\,|{+}{-}\rangle + |{-}{+}\rangle\bigr) + c\,|{-}{-}\rangle,\qquad r,c \in \mathbb R,
\end{equation}
which reduces to $|\psi\rangle_r$ at $c=0$. The corresponding second SRE is
\begin{equation}
\label{M2rc}
M_2(|\psi\rangle_{r,c}) \;=\; \ln \frac{(1+r^2)^4}{1 + 14 r^4 + r^8 + f_c(r)},
\end{equation}
where $f_c(r)$ is a polynomial in $c^2$ vanishing at $c=0$,
\begin{equation}
\label{eq:fc}
\begin{aligned}
f_c(r)=\;&-4c^2\bigl(1+14r^4+r^8\bigr)
+4c^4\bigl(5+7r^2+28r^4+7r^6+5r^8\bigr)\\
&-8c^6\bigl(4+7r^2+14r^4+7r^6+4r^8\bigr)
+16c^8\bigl(1+r^2+r^4\bigr)^2.
\end{aligned}
\end{equation}
The leading correction is negative in $c^2$, so switching on a small $|{-}{-}\rangle$ admixture enhances $M_2$ for fixed $r$. Figure~\ref{fig:M2rc} plots $M_2(|\psi\rangle_{r,c})$ for several fixed values of $c$. Even a modest admixture, $c \lesssim 0.4$, produces magic substantially exceeding the two-coefficient maximum~\eqref{M2rmax}. 

As we will see in the gluon case, $2 \to 2$ gluon scattering with initial helicity $|{+}{-}\rangle$ produces states similar to those in~\eqref{psir}, while $3 \to 2$ scattering produces similar states as~\eqref{psirc}. The finding above therefore motivates us to expect an increase of magic when going from the former case to the latter case.

\subsection{Maximal magic in two-qubit systems} For the most general (complex-coefficient) two-qubit state, the maximum second SRE is conjectured to be~\cite{Liu:2025frx}
\begin{equation}
\label{maxM2bound}
\max_{|\psi\rangle \,\in\, \mathcal{H}_2 \otimes \mathcal{H}_2} M_2(|\psi\rangle) \;=\; \ln \frac{16}{7} \;\approx\; 0.827,
\end{equation}
saturated, for example, by
\begin{equation}
\label{maxM2}
|\psi\rangle_\star \;=\; \frac{i}{2}\,|{+}{-}\rangle - \frac{i}{2}\,|{-}{+}\rangle + \frac{1+i}{2}\,|{-}{-}\rangle.
\end{equation}
Like $|\psi\rangle_{r,c}$ (\ref{psirc}), $|\psi\rangle_\star$ is a superposition of three computational-basis states; unlike $|\psi\rangle_{r,c}$, its coefficients are genuinely complex. The complex phases are essential: the real-coefficient submanifold
$|\psi\rangle_{r,c}$ alone cannot reach the bound~\eqref{maxM2bound}. Its
saturation requires relative phases between basis states. This will matter in Section~\ref{sec:five-point}: the $3 \to 2$ scattering states we construct inherit complex coefficients automatically from the Parke--Taylor and anti-MHV formulas, and we will see that this is what allows them to approach Eq.~\eqref{maxM2bound}.
\section{Basis dependence, phase ambiguity and its cure}
\label{sec:phase}
Although many standard quantities, such as entanglement entropies, are insensitive to local unitary transformations, magic measures depend on the basis for each qubit. To illustrate this, let us first take the state $|\psi\rangle_1=(|+-\rangle+|-+\rangle)/\sqrt2$ as an example. While it is a Bell state in the current basis, redefining the basis by arbitrary phases
\begin{equation}
\label{U14}
    |\pm\rangle_{1,2}'= e^{i\theta_{1,2}^\pm}|\pm\rangle_{1,2}\ ,
\end{equation} 
we can rewrite the state $|\psi\rangle_1$ as
\begin{equation}
\label{psi1}
|\psi\rangle_1 = \frac{1}{\sqrt2}\left(e^{-i(\theta_1^++\theta_2^-)}|{+}{-}\rangle' + e^{-i(\theta_1^-+\theta_2^+)}|{-}{+}\rangle'\right)\ .
\end{equation}
For the case where $\theta_1^++\theta_2^--\theta_1^--\theta_2^+\in \pi \mathbb{Z}$, it is easy to see that (\ref{psi1}) is still a Bell state in the new basis up to an overall phase. Such a state is a stabilizer state and carries no magic. However, once $\theta_1^++\theta_2^-- \theta_1^--\theta_2^+\notin \pi \mathbb{Z}$, this state is generally not a stabilizer state. Indeed, a direct calculation of $M_2$ for (\ref{psi1}) yields
\begin{equation}
\label{M2psi1}
    M_2(|\psi\rangle_1,\{\theta_{1,2}^\pm\})=\ln\frac{16}{14+2\cos 4(\theta_1^++\theta_2^--\theta_1^--\theta_2^+)}\ ,
\end{equation}
which is zero only when $\theta_1^++\theta_2^-- \theta_1^--\theta_2^+\in \frac{\pi}{2}\mathbb Z$.

Of course, the basis can be fixed by specifying a full Pauli frame. However, physical data often defines the $\sigma_z$-eigenstates $\{|+\rangle,|-\rangle\}$ with the transverse Pauli matrices $\sigma_x$ and $\sigma_y$ undetermined, as occurs for the gluon-scattering processes studied below. Consequently, this leads to a phase ambiguity associated to the basis, allowing one to perform a local diagonal redefinition as in (\ref{U14}).

To eliminate the effect of the phase ambiguity, it is useful to average the SRE over the full $U(1)^{2k}$ group corresponding to the basis redefinitions. In fact, from the definition (\ref{Mq}), it is evident that $M_q$ is invariant under an overall phase shift. Consequently, it is sufficient to average over a $U(1)^k$ subgroup that corresponds to the phase shifts for $|+\rangle$'s relative to $|-\rangle$'s,\footnote{We thank Dongheng Qian for pointing this out.}
\begin{equation}
\label{U1k}
|+\rangle'_i=e^{i\theta_i}|+\rangle_i\ .
\end{equation}
Then, for a $k$-qubit state $|\psi\rangle$, we measure the phase-averaged magic by the quantity
\begin{equation}
\label{bMq}
\overline{M_q}(|\psi\rangle) =\frac{1}{(2\pi)^{k}}\left(\prod_{i=1}^k\int_0^{2\pi}d\theta_i\right)M_q(|\psi\rangle,\{\theta_i\})\ .
\end{equation}

Let us take the state (\ref{psir}) as an example and evaluate this phase-averaged SRE. First, employing the redefinition of basis (\ref{U1k}), we rewrite the state as
\begin{equation}
\label{psirtheta}
|\psi\rangle_r = \frac{1}{\sqrt{1+r^2}}\bigl(r\,e^{-i\theta_1}|{+}{-}\rangle' + e^{-i\theta_2}|{-}{+}\rangle'\bigr)\ .
\end{equation}
Up to an overall phase, the dependence of this state on the phases $\theta_{1,2}$ is totally through the combination $\theta_1-\theta_2$. Thus, we anticipate that $M_q$ depends only on such a phase combination. Indeed, a simple calculation of the second SRE for this state yields
\begin{equation}    
\label{M2psirtheta}M_2\left(|\psi\rangle_r,\{\theta_{1,2}\}\right)\;=\; \ln \frac{(1+r^2)^4}{1 +\left[12+2\cos 4(\theta_1-\theta_2)\right] r^4 + r^8}\ .
\end{equation}
When $r=1$, this result reduces to (\ref{M2psi1}).

The integration (\ref{bMq}) with the integrand (\ref{M2psirtheta}) is hard to compute. However, an approximation can be made due to the following observation. In (\ref{M2psirtheta}), the dependence on $\theta_{1,2}$'s is given by the second term in the square bracket. As $\theta_{1,2}$'s vary, the largest amount this term can change is 4, which is relatively small compared with the first term ``12'' in the square bracket. This suggests that the average of this term provides a good approximation to its contribution to the full expression.\footnote{It is interesting to investigate whether this is true in general. We leave it to future work.}

\begin{figure}
\centering
\includegraphics[scale=0.8]{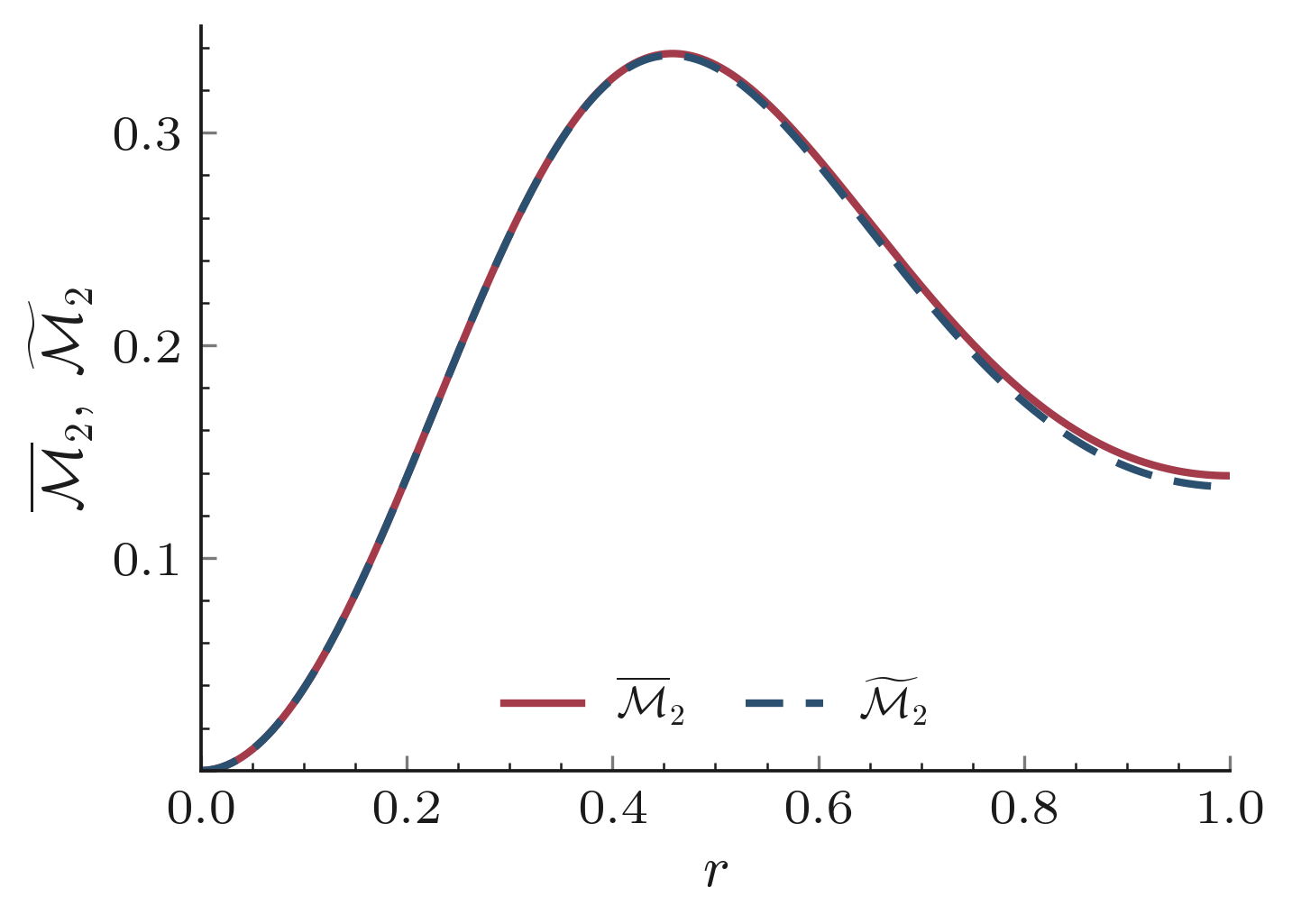}
\caption{\label{fig:bM2r} The second $U(1)^{k}$-invariant stabilizer R\'enyi entropies $\overline{ M_2}$ and $\widetilde{\mathcal{M}}_2$ for the two-coefficient state Eq.~\eqref{psir}, as a function of the amplitude ratio $r$. $\widetilde{\mathcal{M}}_2$ has a global minimum (where it vanishes) at $r=0$, a local minimum at $r=1$, and a global maximum at $r=(\sqrt7-\sqrt3)/2$.}
\end{figure}

Hence, as an approximation to $\overline{M_2}(|\psi\rangle_r,\{\theta_{1,2}\})$, we compute
\begin{equation}    
\label{tM2psirdef}
\widetilde{\mathcal{M}}_2\left(|\psi\rangle_r\right)=-\ln\frac{1}{(2\pi)^2}\int_0^{2\pi}d\theta_1\int_0^{2\pi}d\theta_2\frac{1 +\left[12+2\cos 4(\theta_1-\theta_2)\right] r^4 + r^8}{(1+r^2)^4}\ .
\end{equation}
This integration can be easily performed, yielding
\begin{equation}    
\label{tM2psir}
\widetilde{\mathcal{M}}_2\left(|\psi\rangle_r\right)=\ln \frac{(1+r^2)^4}{1 +12 r^4 + r^8}\ .
\end{equation}
In figure \ref{fig:bM2r}, we plot this result and compare it with the numerically-evaluated $\overline{M_2}$. The two curves are seen to essentially coincide. Moreover, $\widetilde{M}_2$ provides a lower bound on $\overline{M_2}$. This follows directly from Jensen's inequality, since $-\ln x$ is a convex function.

Similarly to (\ref{s0}), setting $\partial \widetilde{\mathcal{M}}_2/\partial r=0$ for (\ref{tM2psir}) gives
\begin{equation}
    r(r^6 - 6r^4 + 6r^2 - 1) \;=\; r(r^2-1)(r^4 - 5r^2 + 1) \;=\; 0.
\end{equation}
The roots in the fundamental region $[0,1]$ are $r=0$, $r=1$ and $r=(\sqrt{7}-\sqrt3)/2$ (see also figure \ref{fig:bM2r}). At $r=0$, corresponding to the global minimum, $\widetilde{\mathcal{M}}_2$ vanishes. This is because $|\psi\rangle_0$ is the trivial state $|{+}{-}\rangle$, which only gains an overall phase for nonzero $\theta_{1,2}$, and thus remains a genuine stabilizer state. The second root $r=1$ corresponds to a local minimum. At this point,
\begin{equation}
\label{tM2ln87}    \widetilde{\mathcal{M}}_2(|\psi\rangle_{r=1})=\ln \frac{8}{7}\approx 0.134\ .
\end{equation}
The nonzero value is due to the fact that this state is no longer a stabilizer state for generic $\theta_{1,2}$, as discussed above. Thus, the degeneracy at $r=0$ and $r=1$, shown in figure \ref{fig:M2r}, has been lifted. Besides, at $r=(\sqrt{7}-\sqrt3)/2$, (\ref{tM2psir}) reaches the maximum, which is
\begin{equation}
\label{tM2ln75}
    \widetilde{\mathcal{M}}_2(|\psi\rangle_{r=\frac{\sqrt7-\sqrt3}{2}})=\ln \frac{7}{5}\approx 0.336\ .
\end{equation}

More generally, we propose a phase-independent SRE, defined as
\begin{equation}
\label{tMq}
    \widetilde{\mathcal{M}}_q(|\psi\rangle) \equiv\frac{1}{1-q}\,\ln\!\left[\frac{1}{(2\pi)^{k}}\left(\prod_{i=1}^k\int_0^{2\pi}d\theta_i\right)\sum_{P \in \mathcal{P}_k} \frac{\langle\psi|P|\psi\rangle^{2q}}{2^k}\right]\ .
\end{equation}
Similar to (\ref{tM2psirdef}), this amounts to an average over $\theta_i$ inside the logarithmic function. Note that an alternative prescription would be to minimize over $U(1)^{k}$ phases, in analogy with the definition of non-local magic~\cite{Cao:2023mzo,Cao:2024nrx,Qian:2025oit}, although in that case the minimization is taken over the full local unitary group. Both prescriptions are meaningful: the minimized one can be used to measure the magic cannot be removed by choosing a phase convention, while the averaged quantity in Eq.~(\ref{tMq}) tells us how much magic is typical over unresolved phase frames.

Before closing this section, we show one more example, which will be useful later for the analysis in Section \ref{sec:five-point}. Consider the state of the form
\begin{equation}
\label{psiBCD0}
|\psi\rangle \;=\;   C\,|{+}{-}\rangle + D\,|{-}{+}\rangle+E\,|{-}{-}\rangle \ ,\quad |C|^2+|D|^2+|E|^2=1\ .
\end{equation}
First, one can compute that the second SRE for this state is
\begin{equation}
\label{M2CDE}
    M_2=-\ln \Big[|C^4+D^4+E^4|^2+12(|CD|^4+|CE|^4+|DE|^4)\Big]\ .
\end{equation}
On the other hand, the redefinition (\ref{U1k}) changes the coefficients of the state as follows:
\begin{equation}
    C\to C e^{-i\theta_1}\ ,\quad D\to D e^{-i\theta_2}\ ,\quad E\to E \ .
\end{equation}
Applying the formula (\ref{tMq}) that integrates out the $\theta_i$-dependence, we obtain the phase-independent second SRE
\begin{equation}
\label{tM2CDE}
    \widetilde{\mathcal{M}}_2=-\ln \left[6\left(|C|^4+|D|^4+|E|^4\right)^2-5\left(|C|^8+|D|^8+|E|^8\right)\right]\ .
\end{equation}
Note that this result depends only on the absolute values $|C|$, $|D|$ and $|E|$ (which are not independent of each other due to the normalization condition in (\ref{psiBCD0})).


\section{Gluons as qubits}
\label{sec:setup}

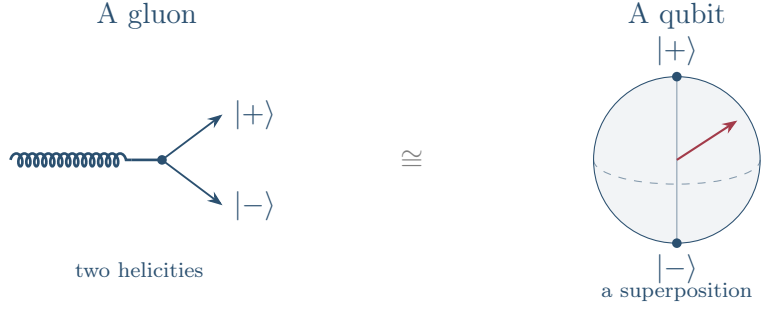
\begin{figure}
\centering
\begin{tikzpicture}[
  >={Stealth[length=2mm]},
  slate/.style={color=LinkSlate},
  garnet/.style={color=UrlGarnet},
  every node/.style={font=\small}
]
\node[slate] at (-3.5,1.9) {A gluon};
\draw[slate, line width=1pt, decorate,
      decoration={coil, aspect=0.6, segment length=3pt, amplitude=2.4pt}]
      (-5.3,0) -- (-3.7,0);
\draw[slate, line width=1pt] (-3.7,0) -- (-3.35,0);
\fill[slate] (-3.3,0) circle (1.8pt);
\draw[slate,->,line width=0.7pt] (-3.3,0) -- (-2.5,0.6) node[right]{$|{+}\rangle$};
\draw[slate,->,line width=0.7pt] (-3.3,0) -- (-2.5,-0.6) node[right]{$|{-}\rangle$};
\node[slate, font=\scriptsize] at (-3.6,-1.45) {two helicities};
\node[gray] at (0,0) {$\cong$};
\node[slate] at (3.5,1.9) {A qubit};
\draw[slate, fill=LinkSlate!6] (3.5,0) circle (1.1);
\draw[slate, dashed, line width=0.4pt, opacity=0.55] (2.4,0) arc (180:360:1.1 and 0.32);
\draw[slate, line width=0.4pt, opacity=0.55] (3.5,1.1) -- (3.5,-1.1);
\fill[slate] (3.5,1.1) circle (1.8pt) node[above]{$|{+}\rangle$};
\fill[slate] (3.5,-1.1) circle (1.8pt) node[below]{$|{-}\rangle$};
\draw[garnet,->,line width=0.8pt] (3.5,0) -- (4.3,0.52);
\node[slate, font=\scriptsize] at (3.5,-1.75) {a superposition};
\end{tikzpicture}
\caption{\label{fig:gluon-qubit}%
A gluon's two helicity states $|{+}\rangle$ and $|{-}\rangle$ furnish a single qubit, and the helicity amplitude is the wavefunction the scattering prepares. A $k$-gluon final state is correspondingly a $k$-qubit state.}
\end{figure}

The central object of this work is the polarization wavefunction produced by tree-level gluon scattering. For each external leg $i$, a gluon is specified by its polarization $\epsilon_i$, momentum $p_i$ and adjoint color index $a_i$. The full single-particle Hilbert space factorises as
\begin{equation}
\label{eq:Hfact}
\mathcal{H} \;\cong\; \mathcal{H}_{\rm pol} \otimes \mathcal{H}_{\rm col} \otimes \mathcal{H}_{\rm kin}.
\end{equation}
A massless gluon has two helicity states, so the polarization sector of a $k$-gluon system is naturally a $k$-qubit Hilbert space (see figure \ref{fig:gluon-qubit}). We use the helicity basis $\{|+\rangle, |-\rangle\}$ on each leg.


It is important to note that the helicity states are defined only up to phases. As we saw in the last section, this phase choice is not innocuous. Specifically, the stabilizer R\'enyi entropies we compute, while Clifford-invariant, are not invariant under arbitrary local unitaries, including local $U(1)$ rephasings. The phase-independent SRE defined in Eq.~(\ref{tMq}) is therefore useful.

Following~\cite{Gargalionis:2025iqs}, we study the gluon state produced by gluon scattering from a certain incoming state of $m$ gluons, and extract the polarization wavefunction of $k$ outgoing gluons by fixing the external colors and momenta. This polarization wavefunction can be read off from the color-dressed amplitude of $m\to k$ scattering, which therefore has $n=m+k$ legs. Throughout we adopt the \emph{all-outgoing} convention: every label $h_i$ denotes the helicity of leg $i$ when the momentum is treated as outgoing. For an outgoing leg, $h_i$ is the physical helicity; for an incoming leg, $h_i$ is the negative of the physical helicity.

In terms of the scattering amplitudes $\mathcal{A}_n$, the polarization wavefunction for the outgoing state can be written as
\begin{equation}
\label{psiA}
|\psi\rangle
\;=\;
\frac{1}{\mathcal{N}}\sum_{h_{m+1},\cdots,h_n=\pm} \mathcal{A}_n(h_1, \ldots, h_m,h_{m+1},\cdots, h_n)\, |h_{m+1} \cdots h_n\rangle,
\end{equation}
where $\mathcal{N}$ denotes the normalization factor. In this paper, we mainly study the two-qubit case, for which
\begin{equation}
\label{psiBCDE}
|\psi\rangle \;=\; B\,|{+}{+}\rangle + C\,|{+}{-}\rangle + D\,|{-}{+}\rangle + E\,|{-}{-}\rangle .
\end{equation}
If there are $m$ incoming gluons with helicities $h_1,\ldots,h_m$, then
\begin{equation}
\label{eq:BCDE}
\begin{aligned}
B&=\mathcal{N}^{-1}\,\mathcal{A}_{m+2}(h_1\cdots h_m\,{+}{+}), &
C&=\mathcal{N}^{-1}\,\mathcal{A}_{m+2}(h_1\cdots h_m\,{+}{-}),\\
D&=\mathcal{N}^{-1}\,\mathcal{A}_{m+2}(h_1\cdots h_m\,{-}{+}), &
E&=\mathcal{N}^{-1}\,\mathcal{A}_{m+2}(h_1\cdots h_m\,{-}{-}).
\end{aligned}
\end{equation}

The $n$-point amplitude admits the Del Duca--Dixon--Maltoni (DDM) decomposition~\cite{DelDuca:1999iql,DelDuca:1999rs},
\begin{equation}
\label{eq:DDM}
\begin{split}
    \mathcal{A}_n\bigl(\{h_i, a_i, p_i\}\bigr)
\;=\;
g^{n-2}\sum_{\sigma \in S_{n-2}} &c[a_1,\sigma(a_2,\cdots,a_{n-1}),a_n]\\
\times &A_n[1, \sigma(2,\ldots,n-1), n]\bigl(\{h_i, p_i\}\bigr),
\end{split}
\end{equation}
where the color factors $c[\sigma]$ are products of structure constants $f^{abc}$,
\begin{equation}
\label{cai}
c[a_1,a_2,\cdots,a_n]=f^{a_1a_2b_1}\left(\prod_{i=1}^{n-4}f^{b_ia_{i+2}b_{i+1}}\right)f^{b_{n-3}a_{n-1}a_n}\ ,
\end{equation}
and the color-ordered partial amplitudes $A_n[\sigma]$ depend only on helicities and momenta. Fixing $\{a_i\}$ turns each $c[\sigma]$ into a c-number, so the resulting amplitude is a function of helicities and momenta. 

A central peculiarity of tree-level gluon amplitudes is that most helicity assignments vanish. If $n_-$ denotes the total number of negative-helicity legs in the all-outgoing convention, then for $n \geq 4$
\begin{equation}
\label{eq:MHVrule}
\mathcal{A}^{\rm tree}_n(\{h_i\}) \;=\; 0
\qquad
\text{whenever } n_- \leq 1 \text{ or } n_- \geq n-1,
\end{equation}
so a tree-level amplitude cannot fire unless at least two helicities differ from the rest.\footnote{Three-point amplitudes are a peculiar exception: for real momenta they vanish by momentum conservation and on-shellness, yet acquire non-trivial values once analytic continuation to complex momenta is permitted. They do not enter the present analysis directly.}

The first non-vanishing class, with $n_- = 2$, is the \emph{maximally helicity-violating} (MHV) sector; its conjugate, with $n_- = n - 2$, is anti-MHV. Within the MHV sector, up to the helicity phase ambiguity mentioned above, the color-ordered partial amplitudes on the r.h.s. of~\eqref{eq:DDM} take the compact Parke--Taylor form~\cite{Parke:1986gb,Nair:1988bq,Mangano:1990by},
\begin{equation}
\label{eq:ParkeTaylor}
A_n[1,2,\cdots,n]^{\rm tree}_{\rm MHV}
\;=\;
\,\frac{\langle ij\rangle^4}{\langle 12\rangle\langle 23\rangle\cdots \langle (n-1)n\rangle \langle n1\rangle},
\end{equation}
where $i$ and $j$ are the two negative-helicity legs. The anti-MHV expression follows by $\langle\,\rangle \to [\,]$. Here, 
\begin{equation}
\label{spinorprod}
\begin{split}
    \langle jl \rangle=[lj]^*=\frac{p_j^1p_l^+-p_j^+p_l^1+i(p_j^2p_l^+-p_j^+p_l^2)}{\sqrt{ |p_j^+p_l^+|}}\ ,\quad p_i^+\equiv p_i^0+ p_i^3
\end{split}
\end{equation}
are the ``square roots''  of the inner product $2|p_j\cdot p_l|$ (also known as the ``spinor inner products'') in the sense that $|\langle jl\rangle|^2=|[lj]|^2=2|p_j\cdot p_l|$.\footnote{Throughout the paper we use the signature where $\eta_{\mu\nu}={\rm diag}\{ +1,-1,-1,-1\}$.}

Taking $m=2$ as an example, there are two qualitatively distinct incoming helicity patterns, represented by $|{-}{-}\rangle$ and $|{+}{-}\rangle$. For $|{-}{-}\rangle$, the MHV selection rule allows only the $|{+}{+}\rangle$ component in the final state, so the two-qubit state is fixed by selection alone and is a stabilizer state with zero magic. For $|{+}{-}\rangle$, two helicity channels survive and the final state is the two-coefficient family
\begin{equation}
\label{psiCD}
|\psi\rangle \;=\; C\,|{+}{-}\rangle + D\,|{-}{+}\rangle ,
\end{equation}
similar to the previous example (\ref{psir}) except that the coefficients are now complex.

More interesting states appear once $m$ is increased. For $m=3$, the incoming state $|{+}{+}{-}\rangle$ produces a three-coefficient family
\begin{equation}
\label{eq:psiBCD}
|\psi\rangle \;=\;  C\,|{+}{-}\rangle + D\,|{-}{+}\rangle+E\,|{-}{-}\rangle .
\end{equation}
This state is analogous to the example (\ref{psirc}) discussed above, but with complex coefficients. In Section \ref{sec:measure}, we showed that even a simple class of three-coefficient states of the form~\eqref{psirc} can exhibit substantially higher magic than the two-coefficient states in~\eqref{psir}. This suggests that a similar enhancement should occur for the more general family in~\eqref{eq:psiBCD}. Indeed, as we will see in Section~\ref{sec:five-point}, the additional $|{-}{-}\rangle$ component is precisely what allows $M_2$ to approach the conjectured two-qubit upper bound, increasing the maximum accessible magic by nearly a factor of three over the four-point case.

\section{Gluons as a source of magic: three examples, three lessons}
\label{sec:examples}

The construction is best met through examples, and gluon scattering, one of the cornerstones of simplicity in the modern scattering amplitude program, supplies a
natural sequence of them. We climb in multiplicity, and the physics changes
character at each rung. The $2\to2$ process exposes the very problem the
phase-independent measure was built to solve: its bare magic looks substantial
but is associated with a phase ambiguity, and the invariant sees through it. The $3\to2$
process delivers the payoff: phase-independent magic that never vanishes and
presses close to the two-qubit maximum. The $2\to3$ process, our first
three-qubit case, supplies the twist: more particles do not buy more magic, and
the invariant settles instead onto a nonzero floor. Through all three runs a
single recurring surprise: the color dependence cancels, leaving the phase-independent magic fixed
by the kinematics alone.

\subsection{Ramp: $2 \to 2$ }
\label{sec:four-point}
In this section, we review the SRE of the gluon state produced by $2\to 2$ scattering~\cite{Gargalionis:2025iqs}, and apply the phase-independent promotion (\ref{tMq}) to it. As discussed above, we focus on the initial helicity state $|{+}{-}\rangle$, which can lead to nontrivial final states,\footnote{More generally, one can start with an initial state that is a superposition of different helicity configurations. See, for example, \cite{Gargalionis:2025iqs}.}
\begin{equation}
\label{psiCD1}
|\psi\rangle \;=\; C\,|{+}{-}\rangle + D\,|{-}{+}\rangle .
\end{equation}
As shown in (\ref{eq:BCDE}), the two nonzero coefficients can be derived by computing the four-point amplitudes $\mathcal A_4({+}{-}{+}{-})$ and $\mathcal A_4({+}{-}{-}{+})$. Specifically, plugging (\ref{eq:ParkeTaylor}) in (\ref{eq:DDM}) gives
\begin{equation}
\label{CDp}
    C=\frac{\langle 24\rangle^4}{\sqrt{\langle 24\rangle^8+\langle 23\rangle^8}}\ ,\quad D=\frac{\langle 23\rangle^4}{\sqrt{\langle 24\rangle^8+\langle 23\rangle^8}}\ ,
\end{equation}
up to the phase ambiguity. Remarkably, the color dependence in $\mathcal A_4$ drops out after normalization by $\mathcal{N}$, because $C$ and $D$ share the same (MHV) Parke--Taylor permutation sum and differ only in their numerators.

It is convenient to work in the center-of-mass frame and parametrize the momenta $p_i$ in terms of the center-of-mass energy $E_{\rm CM}$ and the scattering angle $\theta$ as

\begin{equation}
\begin{aligned}
p_1&=-E_{\rm CM}\,(1,0,0,1)\T, & p_2&=-E_{\rm CM}\,(1,0,0,-1)\T,\\
p_3&=\phantom{-}E_{\rm CM}\,(1,\sin\theta,0,\cos\theta)\T, & p_4&=\phantom{-}E_{\rm CM}\,(1,-\sin\theta,0,-\cos\theta)\T.
\end{aligned}
\end{equation}
Substituting these momenta into (\ref{spinorprod}), one finds
\begin{equation}
    \langle 24\rangle=i\frac{\sqrt2E_{\rm CM}\sin\theta}{\sqrt{1+\cos\theta}}\ ,\quad \langle 23\rangle=-i\frac{\sqrt2E_{\rm CM}\sin\theta}{\sqrt{1-\cos\theta}}\ .
\end{equation}
With these, (\ref{CDp}) can be rewritten as
\begin{equation}
\label{CDtheta}
    C=\frac{(1-\cos\theta)^2}{\sqrt{(1-\cos\theta)^4+(1+\cos\theta)^4}}\ ,\quad D=\frac{(1+\cos\theta)^2}{\sqrt{(1-\cos\theta)^4+(1+\cos\theta)^4}}\ ,
\end{equation}
which depend only on the scattering angle.

Let us temporarily ignore the phase ambiguity and take the coefficients $C$ and $D$ to be positive. The state (\ref{psiCD1}) (\ref{CDtheta}) then takes the form (\ref{psir}), with ratio
\begin{equation}
\label{rtheta}
    r=\frac{(1-\cos\theta)^2}{(1+\cos\theta)^2}=\tan^4\frac{\theta}{2}\ .
\end{equation}
As $\theta$ varies from 0 to $\frac{\pi}{2}$, $r$ ranges from 0 to 1, which is the fundamental region for the state (\ref{psir}) as we discussed before. In particular, at $\theta=0$ the state is trivial, while at $\theta=\frac{\pi}{2}$ it is a Bell state. Using the formula (\ref{M2r}), one finds that the second SRE is
\begin{equation}
\label{M2to2}
    M_2=\ln \frac{(1+\tan^8 \frac{\theta}{2})^4}{1+14 \tan^{16} \frac{\theta}{2}+\tan^{32} \frac{\theta}{2}}\ ,
\end{equation}
in agreement with~\cite{Gargalionis:2025iqs}. 
\begin{figure}
\centering
\includegraphics[scale=0.8]{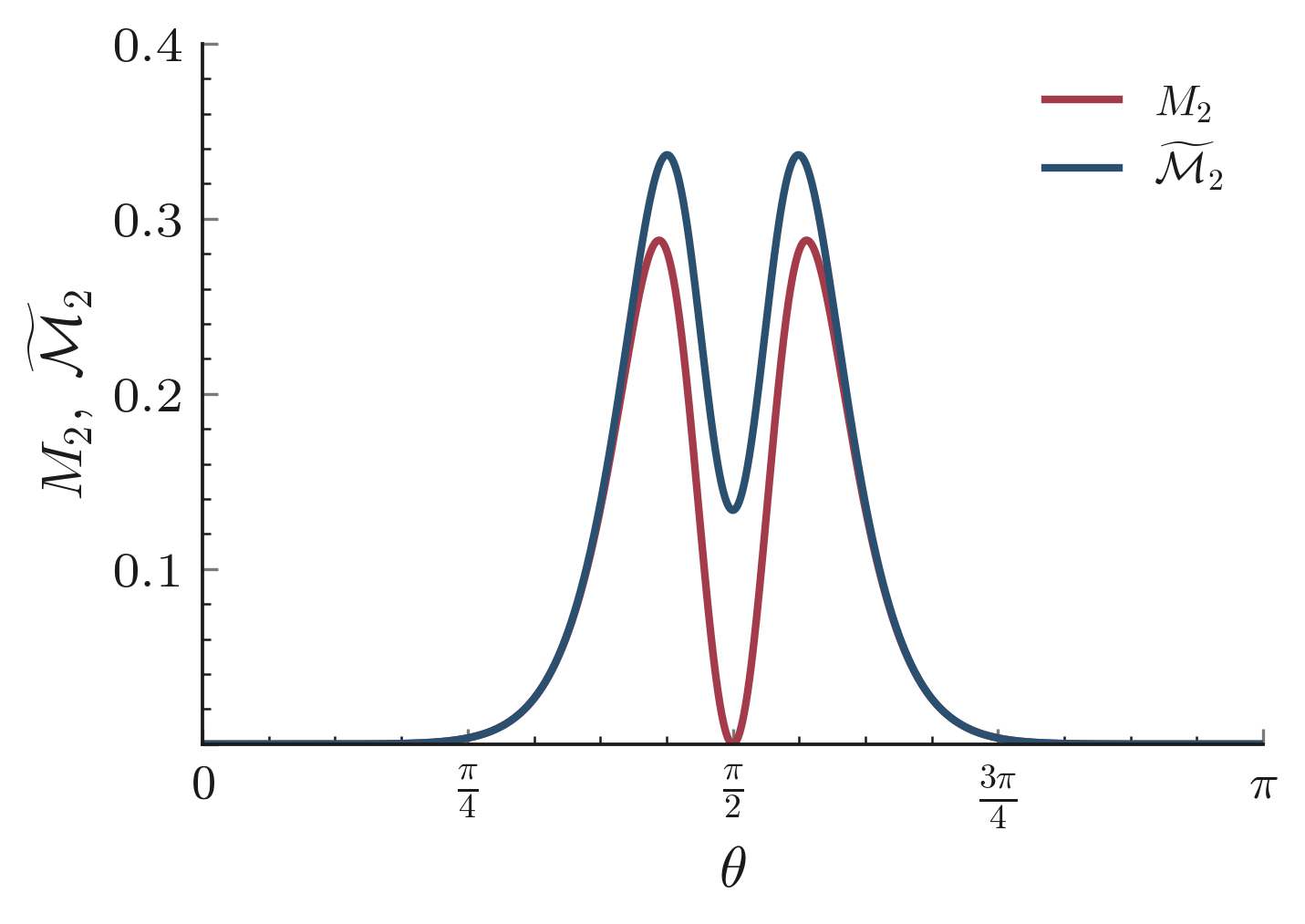}
\caption{\label{fig:M2theta} $M_2$ for $C,D>0$ and $\widetilde{ \mathcal{M}}_2$ as functions of the scattering angle $\theta$ for the final state of $2\to 2$ gluon scattering with initial state $|{+}{-}\rangle$.}
\end{figure}

Once the phase ambiguity in (\ref{CDtheta}) is taken into account, it is more useful to calculate the phase-independent second SRE. To this end, we plug (\ref{rtheta}) in (\ref{tM2psir}) and obtain
\begin{equation}
\label{tM2to2}
  \widetilde{ \mathcal{M}}_2=\ln \frac{(1+\tan^8 \frac{\theta}{2})^4}{1+12 \tan^{16} \frac{\theta}{2}+\tan^{32} \frac{\theta}{2}}\ .
\end{equation}
This result is plotted in figure \ref{fig:M2theta}, together with $M_2$ (\ref{M2to2}). Using $\widetilde{ \mathcal{M}}_2$, we see that the final state never contains magic at scattering angles $\theta=0$ and $\pi$. At $\theta=\frac{\pi}{2}$, $\widetilde{ \mathcal{M}}_2$ reaches a local minimum, (\ref{tM2ln87}). Its maxima occur at $\theta=2\arctan\, (\frac{\sqrt7-\sqrt3}{2})^{1/4}\approx 1.38$ and $\pi-2\arctan\, (\frac{\sqrt7-\sqrt3}{2})^{1/4}\approx 1.77$, where it takes the value in (\ref{tM2ln75}).

\subsection{Substance: $3 \to 2$ }
\label{sec:five-point}
We now proceed to the $3\to 2$ case. Two qualitatively different initial helicity states are $|{-}{-}{-}\rangle$ and $|{+}{+}{-}\rangle$. Due to the MHV selection rule, the former only produces $|{+}{+}\rangle$ as the final state, while the latter results in three possibilities: $|{+}{-}\rangle$, $|{-}{+}\rangle$ and $|{-}{-}\rangle$, and thus the final state is of the form
\begin{equation}
\label{psiBCD}
    |\psi\rangle=  C\,|{+}{-}\rangle + D\,|{-}{+}\rangle+E\,|{-}{-}\rangle\ .
\end{equation}
Our focus in this section will be on this nontrivial case.

As shown in (\ref{eq:BCDE}), the nonzero coefficients in (\ref{psiBCD}) are given by the following five-point amplitudes
\begin{equation}
    C=\mathcal N^{-1} \mathcal A_5({+}{+}{-}{+}{-}), D=\mathcal N^{-1} \mathcal A_5({+}{+}{-}{-}{+}),  E=\mathcal N^{-1} \mathcal A_5({+}{+}{-}{-}{-})\ .
\end{equation}
Combining~\eqref{eq:DDM},~\eqref{cai} and~\eqref{eq:ParkeTaylor}, we can write these coefficients explicitly as
\begin{equation}
\label{BCD12345}
C=\mathcal N^{-1}\,\ab{3}{5}^4\,\Ssum\ ,\qquad
D=\mathcal N^{-1}\,\ab{3}{4}^4\,\Ssum\ ,\qquad
E=\mathcal N^{-1}\,\sb{1}{2}^4\,\Sbar\ ,
\end{equation}
up to the phase ambiguity, where the color-kinematic sum and its conjugate are
\begin{align}
\label{Ssum}
\Ssum&\equiv\sum_{\sigma\in S_{n-2}}\frac{\fff}{\PTden}\ ,\nonumber\\
\Sbar&\equiv\sum_{\sigma\in S_{n-2}}\frac{\fff}{\PTdens}\ .
\end{align}

Compared with the previous section, one complication in the $3\to2$ case might be that while the ratio between $C$ and $D$ is still independent of the color configuration $\{a_i\}$, the ratios $C/E$ and $D/E$ generally have nontrivial color dependence due to the difference between the spinor products $\langle\,\rangle$ and $[\,]$.

However, spinor products have the property that $\ab{i}{j}=-[ij]^*$ (see~\eqref{spinorprod}). Consequently, $\Ssum=-\Sbar^*$, which means that $|\Ssum|=|\Sbar|$. It then follows from~\eqref{BCD12345} that
\begin{equation}
\label{ratioCDE}
    |C|:|D|:|E|=|\langle 35\rangle^4|:|\langle 34\rangle^4|:|[12]^4|\ ,
\end{equation}
which clearly contains no color dependence. Therefore, the phase-independent second SRE (\ref{tM2CDE}) is also insensitive to the color configuration, and takes the form
\begin{equation}
\label{tM2sp}
\begin{split}
   \widetilde{\mathcal{M}}_2
=\ln\frac{\left(\left|\langle 35\rangle^8\right|+\left|\langle 34\rangle^8\right|+\left|[12]^8\right|\right)^4} {6\left(\left|\langle 35\rangle^{16}\right|+\left|\langle 34\rangle^{16}\right|+\left|[12]^{16}\right|\right)^2 -5\left(\left|\langle 35\rangle^{32}\right|+\left|\langle 34\rangle^{32}\right|+\left|[12]^{32}\right|\right)}\ .    
\end{split}
\end{equation}

Another complication for the five-point case lies at the kinematic configuration. In the previous section, we see that there is only one parameter in the final state (\ref{psiCD1}), (\ref{CDtheta}), namely the scattering angle $\theta$. In contrast, the $3\to 2$ gluon scattering is parametrized by more scattering angles (between various two momenta). Furthermore, even in the center-of-mass frame, the three incoming momenta do not necessarily have the same energy, which potentially gives rise to nontrivial energy dependence in the final state.

For simplicity, let us focus on the symmetric configuration in which the three
incoming momenta have equal energy. We parametrize the five external momenta as
\footnotesize
\begin{equation}
\label{pi5}
\begin{aligned}
p_1&=-E_{\rm CM}\,(1,0,0,1)\T,\qquad
p_2=-E_{\rm CM}\left(1,0,\frac{\sqrt3}{2},-\frac12\right)\T,\\
p_3&=-E_{\rm CM}\left(1,0,-\frac{\sqrt3}{2},-\frac12\right)\T,\qquad
p_4=\frac{3E_{\rm CM}}{2}
\left(1,\sin\theta\cos\phi,\sin\theta\sin\phi,\cos\theta\right)\T,\\
p_5&=\frac{3E_{\rm CM}}{2}
\left(1,-\sin\theta\cos\phi,-\sin\theta\sin\phi,-\cos\theta\right)\T .
\end{aligned}
\end{equation}
\normalsize
At $\theta=0$, the outgoing momenta $p_4$ and $p_5$ align with $p_1$ while at $(\theta,\phi)=(\frac{\pi}{2},0)$, they are perpendicular to the plane formed by the incoming momenta.

\begin{figure}
\centering
\includegraphics[scale=0.8]{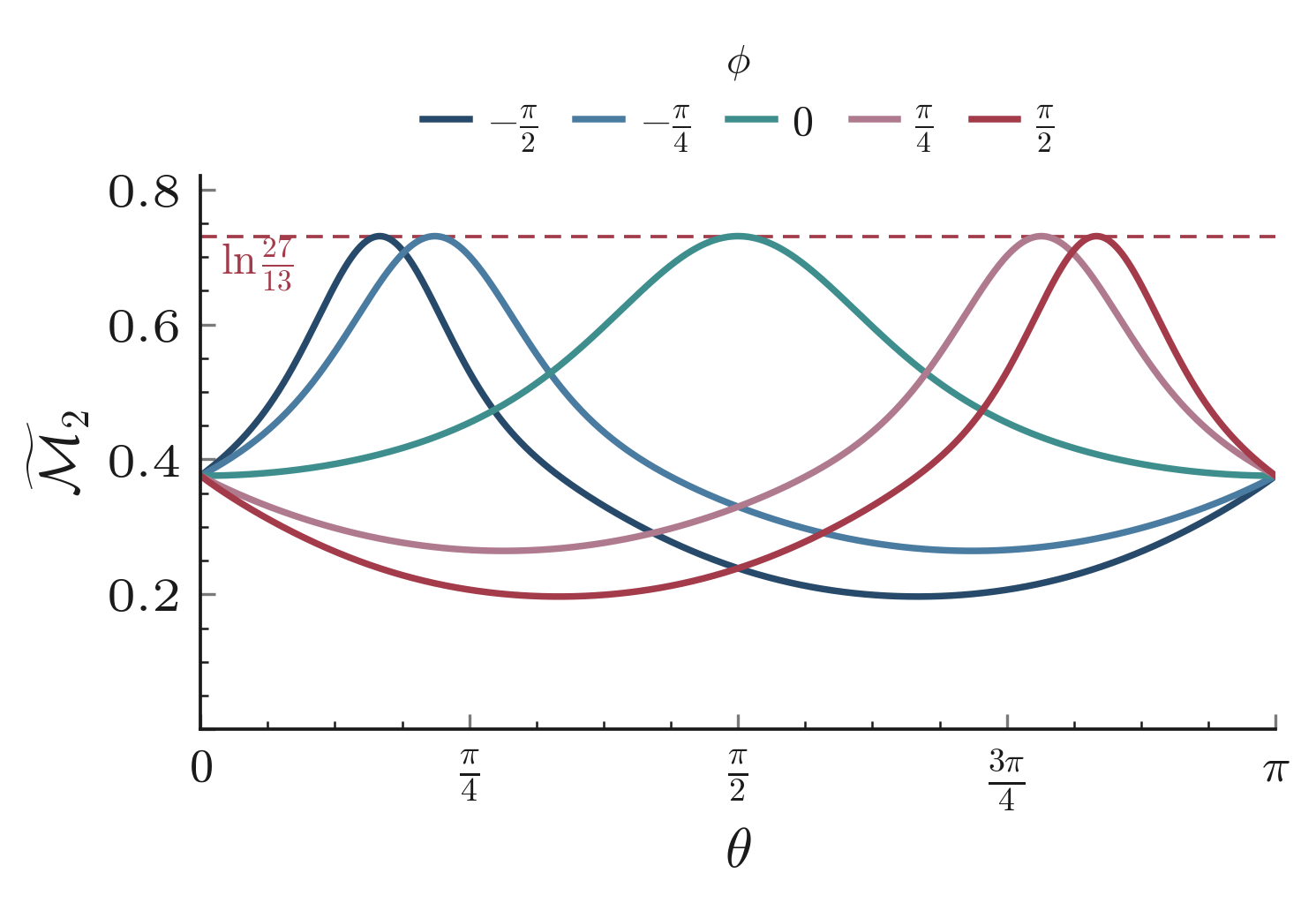}
\caption{\label{fig:tM2phitheta11123} $\widetilde{\mathcal{M}}_2$ for the final state of $3\to 2$ gluon scattering with initial helicity $|{+}{+}{-}\rangle$. The momenta are parametrized as in (\ref{pi5}).}
\end{figure}
As an example, when $(\theta,\phi)=(\frac{\pi}{2},0)$, the phase-independent second SRE (\ref{tM2sp}) $\widetilde{\mathcal{M}}_2=\ln \frac{27}{13}\approx 0.731$, which is significantly larger than the one that any $ 2\to 2$ scattering can achieve (see figure \ref{fig:M2theta}). More generally, we plot $\widetilde{\mathcal{M}}_2$ as a function of $\theta$ for several fixed values of $\phi$ in figure \ref{fig:tM2phitheta11123}. It shows that the value $\widetilde{\mathcal{M}}_2=\ln \frac{27}{13}$ for $(\theta,\phi)=(\frac{\pi}{2},0)$ is the maximum. Furthermore, for any fixed value of $\phi$, $\widetilde{\mathcal{M}}_2$ can reach the same maximal value $\ln \frac{27}{13}$ as $\theta$ varies.

Interestingly, comparing $3\to 2$ scattering (figure \ref{fig:tM2phitheta11123}) with $2\to 2$ (figure \ref{fig:M2theta}) reveals that the former case generally gives rise to higher magic than the latter case, as expected from the simple examples discussed in Section \ref{sec:measure}. Also note that $\widetilde{\mathcal{M}}_2$ vanishes nowhere in the former case.

It is instructive to consider a particular color configuration and evaluate $M_2$ with the phase ambiguity ignored, for comparison with the phase-independent quantity $\widetilde{\mathcal{M}}_2$. To this end, let us pick a simple configuration $\{a_1,a_2,a_3,a_4,a_5\}=\{1,1,1,2,3\}$. In this case, the only possible color factors in (\ref{BCD12345}) are
\begin{equation}
f^{a_1a_4b_1}f^{b_1a_3b_2}f^{b_2a_2a_5}\text{ and } f^{a_1a_4b_1}f^{b_1a_2b_2}f^{b_2a_3a_5}\ ,
\end{equation}
which are equal to each other because $a_2=a_3$. The color-dependent sums (\ref{Ssum}) are then simplified to
\begin{equation}
\label{Ssum12345}
\begin{split}
    \Ssum=&\frac{1}{\langle 14\rangle\langle 43\rangle\langle 32\rangle\langle 25\rangle\langle 51\rangle}+\frac{1}{\langle 14\rangle\langle 42\rangle\langle 23\rangle\langle 35\rangle\langle 51\rangle}\ ,\\
    \Sbar=&\frac{1}{[14][43][32][ 25][51]}+\frac{1}{[14][42][23][ 35][51]}\ .    
\end{split}
\end{equation}
It is noteworthy that when the outgoing momenta are perpendicular to all the incoming momenta, or $(\theta,\phi)=(\frac{\pi}{2},0)$, the state becomes
\begin{equation}
    |\psi\rangle_{\theta=\frac{\pi}{2},\phi=0}=\frac{(1+\sqrt3 i)^2}{4\sqrt3}|{+}{-}\rangle-\frac{1+\sqrt3 i}{2\sqrt3}|{-}{+}\rangle+\frac{1}{\sqrt3}|{-}{-}\rangle\ ,
\end{equation}
up to the phase ambiguity. Its (bare) second SRE (\ref{M2CDE}) is $\ln \frac{9}{4}\approx 0.811$, which is only slightly lower than the (bare) maximum (\ref{maxM2bound}) ($\frac{9}{4}=\frac{63}{28}<\frac{64}{28}=\frac{16}{7}$).

\begin{figure}
\centering
\includegraphics[scale=0.8]{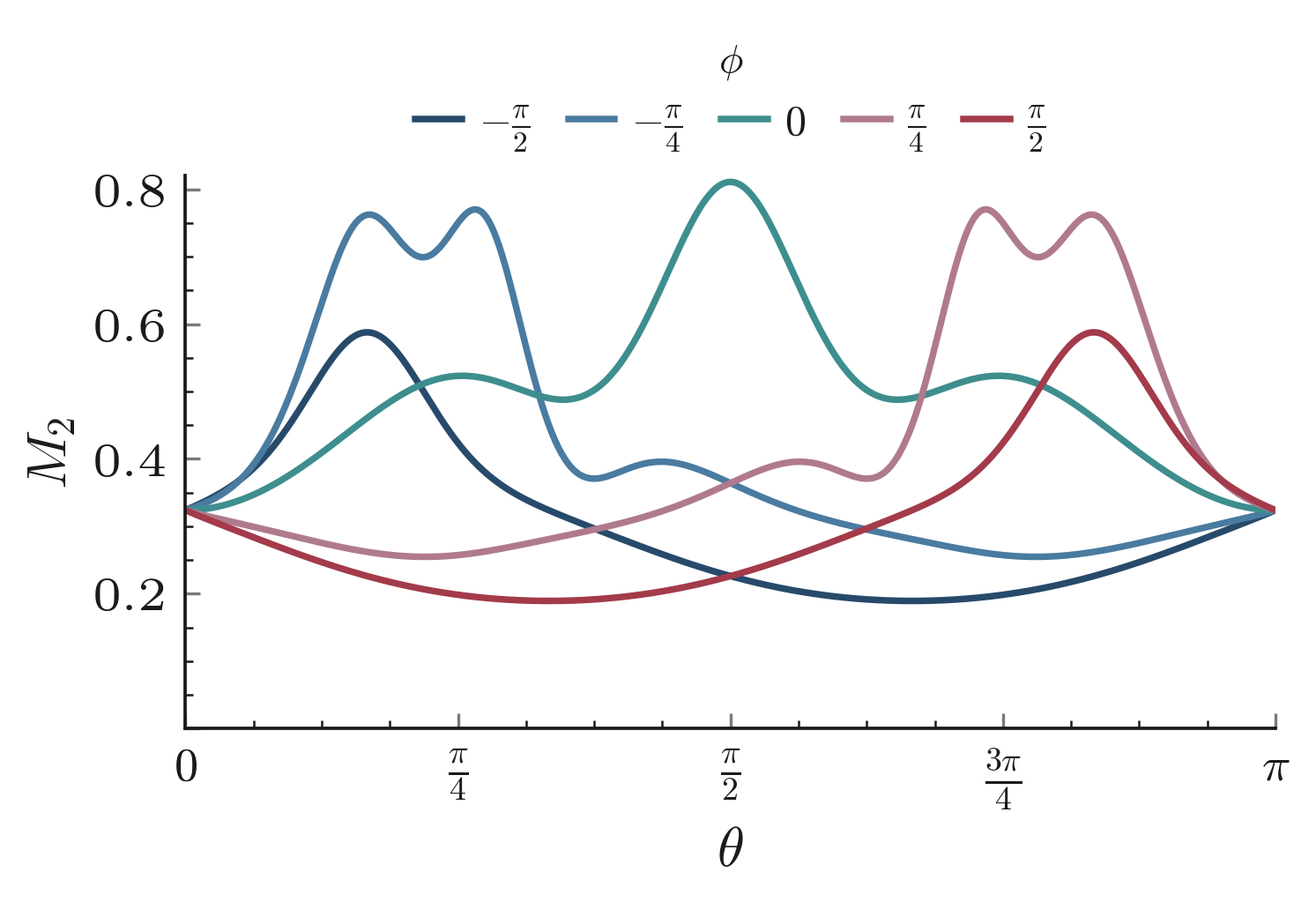}
\caption{\label{fig:M2phitheta11123} $M_2$ for the final state of $3\to 2$ gluon scattering with initial helicity $|{+}{+}{-}\rangle$. The momenta are parametrized as in (\ref{pi5}) and the color configuration is $\{a_1,a_2,a_3,a_4,a_5\}=\{1,1,1,2,3\}$.}
\end{figure}

In figure \ref{fig:M2phitheta11123}, we plot $M_2$ for the state (\ref{psiBCD}) with (\ref{BCD12345}) and (\ref{Ssum12345}). One can see that this figure contains some wiggles, which are artifacts of ignoring the phase ambiguity in (\ref{Ssum12345}). In contrast, for the phase-independent one, plotted in figure \ref{fig:tM2phitheta11123}, these wiggles have been eliminated by the average over phase redefinitions. Besides, figure \ref{fig:M2phitheta11123} shows that the maximal $M_2$ is reached at only the point $(\theta,\phi)=(\frac{\pi}{2},0)$, and thus the ability of reaching the same maximal (phase-independent) magic for different $\phi$ discussed above (exhibited in figure \ref{fig:tM2phitheta11123}) is hidden.

\subsection{Plateau: $2\to 3$}
So far we have focused on the magic for two-qubit states. In this section, we generalize the discussion to three-qubit cases. Specifically, we consider states produced from $2\to 3$ gluon scattering, with the initial helicity state $|{+}{+}\rangle$ for simplicity. Applying the MHV selection rule, the final state takes the form 
\begin{equation}
\label{psi3qb}
    |\psi\rangle=B|{+}{-}{-}\rangle+C|{-}{+}{-}\rangle+D|{-}{-}{+}\rangle+E|{-}{-}{-}\rangle\ .
\end{equation}
One can calculate using (\ref{tMq}) that the corresponding phase-independent second SRE is
\begin{equation}
\label{tM2BCDE}
    \widetilde{\mathcal{M}}_2=-\ln \left[6\left(|B|^4+|C|^4+|D|^4+|E|^4\right)^2 -5\left(|B|^8+|C|^8+|D|^8+|E|^8\right) \right]\ .
\end{equation}
Accidentally, when $B=0$, this result coincides with (\ref{tM2CDE}) for two-qubit states.

As shown in (\ref{psiA}), the coefficients in (\ref{psi3qb}) can be computed from the five-point amplitudes:
\begin{equation}
\begin{split}
    B=\mathcal N^{-1} \mathcal A_5({+}{+}{+}{-}{-})\ ,\quad &
    C=\mathcal N^{-1} \mathcal A_5({+}{+}{-}{+}{-})\ , \\
    D=\mathcal N^{-1} \mathcal A_5({+}{+}{-}{-}{+})\ ,\quad & E=\mathcal N^{-1} \mathcal A_5({+}{+}{-}{-}{-})\ .  
\end{split}
\end{equation}
Since these amplitudes are either in the MHV sector or the anti-MHV sector, they can be computed using the Parke--Taylor form. Specifically, plugging (\ref{eq:ParkeTaylor}) in (\ref{eq:DDM}) gives
\begin{equation}
\label{BCDE12345}
B=\mathcal N^{-1}\,\ab{4}{5}^4\,\Ssum\, ,\;
C=\mathcal N^{-1}\,\ab{3}{5}^4\,\Ssum\, ,\;
D=\mathcal N^{-1}\,\ab{3}{4}^4\,\Ssum\, ,\;
E=\mathcal N^{-1}\,\sb{1}{2}^4\,\Sbar\, .
\end{equation}
These share the same sum $\Ssum$ as the $3\to2$ case~\eqref{BCD12345}. Furthermore, similar to (\ref{ratioCDE}), the ratio among the absolute values of the coefficients is simply
\begin{equation}
    |B|:|C|:|D|:|E|=|\langle 45\rangle^4|:|\langle 35\rangle^4|:|\langle 34\rangle^4|:|[12]^4|\ ,
\end{equation}
which is independent of the color factors. As a result, the expression of $\widetilde{\mathcal{M}}_2$ shown in (\ref{tM2BCDE}), which depends only on the absolute values of the coefficients, can be expressed as
\footnotesize
\begin{equation}
\label{tM2BCDEsp}
    \widetilde{\mathcal{M}}_2=\ln { \frac{\left(\left|\langle 45\rangle^8\right|+\left|\langle 35\rangle^8\right|+\left|\langle 34\rangle^8\right|+\left|[12]^8\right|\right)^4}{6\left(\left|\langle 45\rangle^{16}\right|+\left|\langle 35\rangle^{16}\right|+\left|\langle 34\rangle^{16}\right|+\left|[12]^{16}\right|\right)^2 -5\left(\left|\langle 45\rangle^{32}\right|+\left|\langle 35\rangle^{32}\right|+\left|\langle 34\rangle^{32}\right|+\left|[12]^{32}\right|\right)}}\ ,
\end{equation}
\normalsize
regardless of the color configurations.

Remarkably, one can see in (\ref{tM2BCDEsp}) that $\widetilde{\mathcal{M}}_2$ depends on the incoming momenta $p_1$ and $p_2$ only through $|[ 12]|$. Furthermore, in the center-of-mass frame, $|[12]|=E_{\rm CM}$, where the center-of-mass energy $E_{\rm CM}=-2(p_1)^0=-2(p_2)^0$ (recall that our convention is such that the incoming energy is negative). Hence, $\widetilde{\mathcal{M}}_2$ is independent of the spatial momenta of the two incoming gluons. This enables us to parametrize the outgoing momenta as
\begin{equation}
\label{p345}
\begin{split}
    p_3=&E_{\rm CM1}(1,0,0,1)\T\ ,\quad p_4=E_{\rm CM2}(1,0,\sin\theta,\cos\theta)\T\ ,\quad \\
    p_5=&(\sqrt{E_{\rm CM1}^2+E_{\rm CM2}^2+2\cos\theta E_{\rm CM1}E_{\rm CM2}},(-\vec{p}_3-\vec{p}_4)\T)\T    
\end{split}
\end{equation}
without loss of generality. Note that in terms of the kinematic variables $\theta$, $E_{\rm CM1}$ and $E_{\rm CM2}$, the center-of-mass energy can be expressed as
\begin{equation}
    E_{\rm CM}=E_{\rm CM1}+E_{\rm CM2}+\sqrt{E_{\rm CM1}^2+E_{\rm CM2}^2+2E_{\rm CM1}E_{\rm CM2}\cos\theta}\ .
\end{equation}
\begin{figure}
\centering
\includegraphics[scale=0.8]{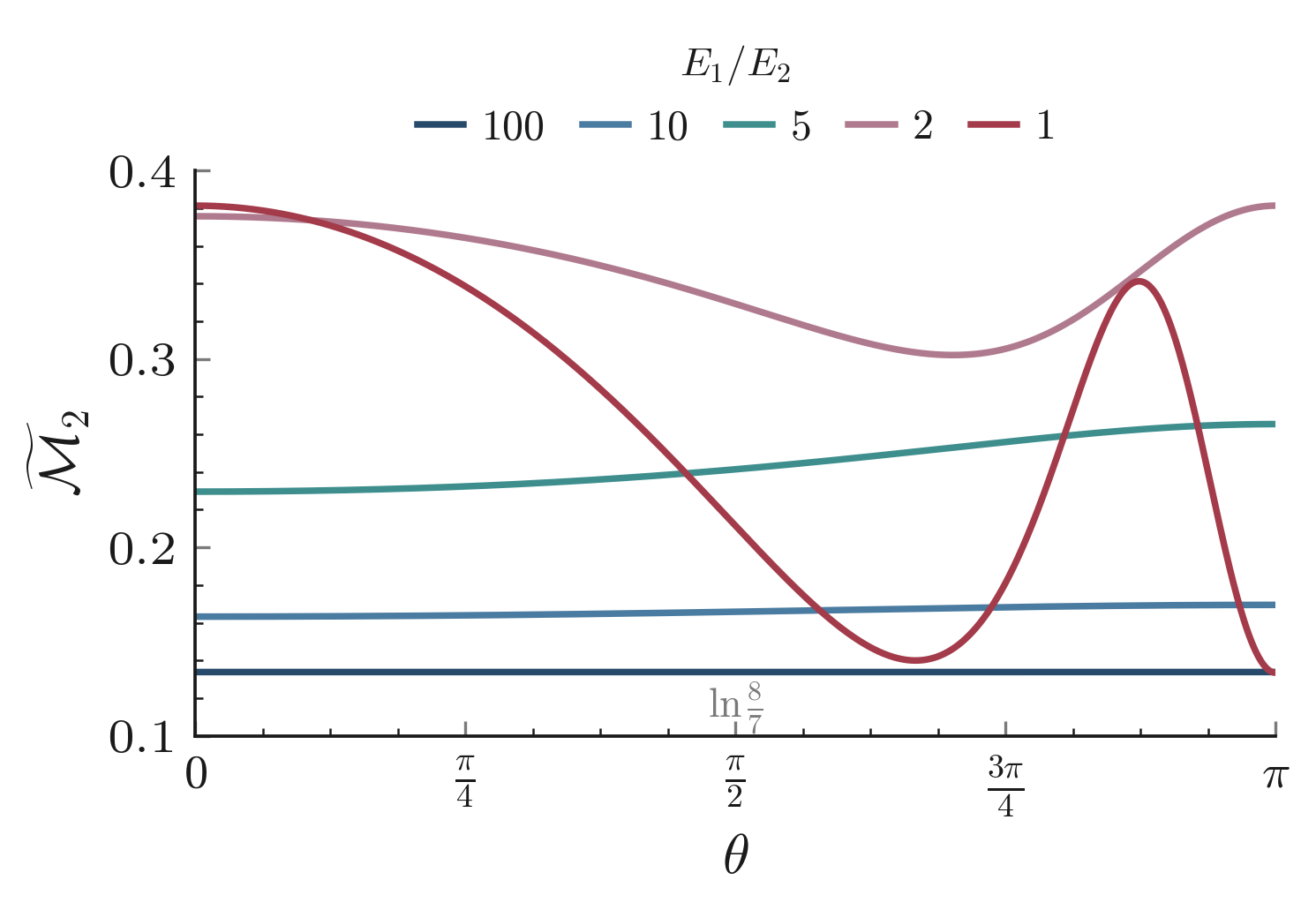}
\caption{\label{fig:tM2E12theta} $\widetilde{\mathcal{M}}_2$ for the final state of $2\to 3$ gluon scattering with initial helicity $|{+}{+}\rangle$. The incoming momenta are parametrized as in (\ref{p345}).}
\end{figure}

In figure \ref{fig:tM2E12theta}, we plot $\widetilde{\mathcal{M}}_2$ as a function of $\theta$ for different fixed energy ratios $E_{\rm CM1}/E_{\rm CM2}$. One can see that the magic is bounded from both above and below. The upper bound, which is about 0.38, is even lower than the $\widetilde{\mathcal{M}}_2$ that most two-qubit states in the previous $3\to 2$ case can achieve (see figure \ref{fig:tM2phitheta11123}). On the other hand, the lower bound, which is about 0.13, can be approached by taking the soft limit, corresponding to the regime $E_{\rm CM2}\ll E_{\rm CM1}$, or the regime $E_{\rm CM2}\to E_{\rm CM1}$ with $\theta\to \pi$ in figure \ref{fig:tM2E12theta}. 

Interestingly, figure \ref{fig:tM2E12theta} also shows that in the limit of $E_{\rm CM1}\to E_{\rm CM2}$ and $\theta\to \frac{2\pi}{3}$, where the three outgoing energies become the same, $\widetilde{\mathcal{M}}_2$ is close to the lower bound set by the soft limit. To compare them, let us evaluate the two values of $\widetilde{\mathcal{M}}_2$ analytically. First, in the soft limit $E_{\rm CM2}/E_{\rm CM1}\to 0$, we have
\begin{equation}
\label{tM2soft}
    |B|:|C|:|D|:|E|=0:1:0:1\Rightarrow \widetilde{\mathcal{M}}_2=\ln\frac{8}{7}\approx 0.134\ .
\end{equation}
On the other hand, in the ``symmetric limit'' $E_{\rm CM1}=E_{\rm CM2}$ and $\theta=\frac{2\pi}{3}$,
\begin{equation}
    |B|:|C|:|D|:|E|=1:1:1:9\Rightarrow \widetilde{\mathcal{M}}_2=\ln\frac{4148928}{3606913}\approx 0.140\ ,
\end{equation}
which is only larger than the lower bound (\ref{tM2soft}) by a tiny amount.

\section{Summary and outlook}
\label{sec:outlook}

Magic is the resource behind quantum advantage, yet the standard measure of it
is not a property of the state: it depends on a choice of basis. On the other hand, physical data often specify the basis up to a phase ambiguity. In this paper, we generalized the notion of magic by averaging the phase-dependent terms in the magic measure stabilizer R\'enyi entropy (SRE). The resulting quantity is thus independent of the basis phase.

Applied to gluons, this strategy tells a clean story. 
For $2\to2$ scattering, we found that as the scattering angle varies, the phase-independent second SRE $\widetilde{M}_2$ ranges from zero to $\ln (7/5)\approx 0.34$. We then push beyond the familiar four-point setting to higher-point gluon amplitudes, where a richer pattern emerges. At $3\to2$ the magic never vanishes and is generically larger than the maximum that $2\to 2$ can achieve. At $2\to3$ magic refuses to grow:
more outgoing legs lower the ceiling, and $\widetilde{M}_2$ exhibits a floor $\ln(8/7)\approx0.13$ that corresponds to the soft limit. Moreover, close to this floor lies a local minimum of approximately 0.14, occurring when the three outgoing momenta are symmetric. In every case we
computed, the color cancels: the phase-independent magic of a gluon state is carried
by the helicities alone.

Four questions follow, and we leave them open. First, we do not know how far the color
cancellation reaches: whether it is a theorem or an accident of low multiplicity. Second,
we do not know whether $\widetilde{\mathcal{M}}_q$ remains a resource monotone, as the (bare)
SRE is for $q\geq2$. Besides, as mentioned before one may average over the phase orbit, as we do, or minimize over it; the two are different phase-independent quantities, and which is the right one is not settled.

A fourth question is experimental. The moduli that determine $\widetilde{\mathcal{M}}_2$ are helicity-amplitude moduli, in principle the target of polarization tomography, and the phases we average over are precisely what such data leave unresolved; this is the physical motivation for the average. But color decoherence, hadronization, and the difficulty of gluon polarimetry stand between our tree-level diagnostic and any collider measurement, in contrast to the top-quark case~\cite{ATLAS:2023fsd,CMS:2025cim}. It is therefore an interesting direction for future work to explore how the phase-independent stabilizer Rényi entropy introduced here can be connected to experimentally accessible observables.

We emphasize that none of these is special to gluons. Gluons are the most studied case of the modern amplitudes
program; gravitons, fermions, the double copy, loops, the string, each prepares
its own wavefunction, and now has a magic of its
own. Entanglement showed that amplitudes are quantum information in disguise.
Magic asks the sharper question: not how correlated the $S$-matrix makes the
world, but how hard it is to fake. The gluon's answer already surprised us and we are scratching the surface. It
would be surprising if there is not more magic in scattering amplitudes.

\acknowledgments
SK gratefully acknowledges CERN for its invitation and hospitality, and in particular the TH Institute on Quantum Simulation and Computation in High-Energy Physics, where this project originated. We also thank Qiyuan Hu, Sam Li, and Dongheng Qian for valuable discussions, careful reading of the draft, and insightful feedback.

\bibliographystyle{utphys}
\bibliography{paper}{}

\end{document}